\documentclass[11pt,a4paper]{article}
\usepackage{a4}
\title{Quantum Coherence in a Single Ion due to strong Excitation of a metastable Transition}

\author{J. von Zanthier$^{1}$, C. Skornia$^{1,2}$, G. S. Agarwal$^{1,3}$ and H. Walther$^{1}$ \\ 
$^{1}$ Max-Planck-Institut f\"ur Quantenoptik and\\
Sektion Physik der LMU M\"unchen, D-85748 Garching, Germany \\
$^{2}$ Institut f\"ur Theoretische Physik,\\
Universit\"at Regensburg, D-93040 Regensburg, Germany\\
$^{3}$ Physical Research Laboratory, Navrangpura, Ahmedabad-380 009, India}

\begin{document}
\maketitle

\begin{abstract}

We consider pump-probe spectroscopy of a single ion with a highly metastable (probe) clock transition which is monitored by using the quantum jump technique. For a weak clock laser we obtain the well known Autler-Townes splitting. For stronger powers of the clock laser we demonstrate the transition to a new regime. The two regimes are distinguished by the transition of two complex eigenvalues to purely imaginary ones which can be very different in magnitude. The transition is controlled by the power of the clock laser. For pump on resonance we present simple analytical expressions for various linewidths and line positions. \\

PACS number(s): 42.50.Gy, 42.50.Hz, 32.80.Qk

\end{abstract}

\section{Introduction} \label{Intro}

Ever since the first observation of coherent population trapping (CPT) \cite{Alzetta76} - \cite{Arimondo96} this phenomenon has stimulated experimental and theoretical investigations. Starting with the study of dark resonances in $\Lambda$-systems where two lower levels are coupled to a common upper state by two coherent light fields \cite{Arimondo76} new phenomena like electromagnetically induced transparency (EIT) \cite{Harris91}, enhanced index of refraction without absorption \cite{Scully91} and lasing without inversion (LWI) \cite{Harris89} - \cite{Padmabandu96} have been investigated. Due to the potentially extreemly steep dispersion at the CPT resonance, group velocities were recently detected which are orders of magnitude smaller than the vacuum speed of light \cite{Hau99,Scully99}. 

Coherent population trapping in a $\Lambda$-system arises when a coherence between the two lower levels is established by a nonlinear process. Experimentally this can be achieved by applying two lasers simultaneously, each coupling one of the lower levels near resonantly to the common upper state. When the difference between the two laser frequencies matches exactly the lower level splitting (which can be zero) the atom is optically pumped into a coherent superposition of the two lower states. In this case no further light is absorbed \cite{Arimondo96}. Since the two ground states are radiatively stable, the ground-state coherence can have long lifetimes leading to extremely narrow resonances.

Systems with other level schemes have also been studied, e.g. V-systems or cascade-systems. Needless to say that many of the properties of these systems have been investigated experimentally and theoretically in much detail \cite{Arimondo96}. In this work we will focus on the discussion of nonlinear behaviour in a V-system, where the two upper states have very {\it different lifetimes}. 

A theoretical investigation of this kind is triggered by our work with single In$^+$ ions \cite{Becker00} where the V-system is realized by the $5s^{2 \enspace 1}S_{0}$ ground state coupled to the two upper levels $5s5p \, ^{3}P_{1}$ and $5s5p \, ^{3}P_{0}$, respectively (see Fig. 1c). Since the $5s^{2 \enspace 1}S_{0}$ - $5s5p \, ^{3}P_{0}$ intercombination line (clock transition), induced by magnetic dipole hyperfine interaction, is highly suppressed, the lifetime of this state is rather long, with a lifetime of $\tau ( ^{3}P_{0} ) \simeq$ 195 ms ($\Gamma_{^{3}P_{0}} \simeq 2 \pi \cdot$ 0.82 Hz). It is longer by a factor of  $\sim 10^5$ than the lifetime of the dipole allowed $5s^{2 \enspace 1}S_{0}$ - $5s5p \, ^{3}P_{1}$ (cooling) transition, with $\tau ( ^{3}P_{1} ) \simeq$ 440 ns ($\Gamma_{^{3}P_{1}} \simeq 2 \pi \cdot$ 360 kHz). The {\it narrow} $5s^{2 \enspace 1}S_{0}$ - $5s5p \, ^{3}P_{0}$ transition used as a probe allows thus to explore sensitively unusual features and regimes. 

Single trapped ions are particularly useful for studying coherence effects. They fulfill most prerequisits needed for ultra-high resolution spectroscopy: By cooling the ion to the ground state of the trap the particle is well localized in the Lamb-Dicke-regime \cite{Stenholm86} and almost free of motion. Thus transit-time and first-order Doppler linebroadening are eliminated and the second-order Doppler effect can be reduced to negligible values. Sufficiently low pressures in the vacuum chamber avoid broadening due to collisions. In the case of a single ion no averaging over spacially varying laser intensities or atomic densities has to be taken into account, and no trap dynamics (e.g. spacially varying Zeemanshifts) need to be considered. In case of In$^+$, due to the absence of hyperfine structure and the metastability of level $5s5p \, ^{3}P_{0}$, no repumping or Zeeman-substructure has to be taken into account in the theoretical description. A true closed 3-level system is realized in this case. As far as nonlinear spectroscopy with single ions is concerned coherence effects have been investigated in Ba$^+$ \cite{Toschek81} - \cite{Toschek92}. Here, a $\Lambda$-system was investigated where the laser transitions were strongly dipole-allowed. The participating states had thus similar transition probabilities. This is in contrast to the level scheme investigated here where the two upper states of the V-system have very different lifetimes. This is what leads to the unusual coherence effects investigated in this paper.

The paper is organized as follows: in Section 2 we introduce our theoretical model and present numerical results for the fluorescence on the fast transition. We discuss different regimes of fluorescence depending on the power of the clock laser. In Section 3 we show how the numerical results of Section 2 can be understood in terms of analytical calculations if we specialize for the case of on-resonance pumping. In Section 4 we deal with the general case of arbitrary pump-detuning. Here, only numerical results are presented, as space does not permit us to present the analytical results.

\section{Clock Laser Induced Coherence} \label{}

In order to model the V-configuration in In$^+$ we consider an atomic system with levels $\mid 1 \ \rangle$ and $\mid 2 \ \rangle$ which are coupled to a common lower level $\mid 3 \ \rangle$ (see Fig. 1a). As in In$^+$, we suppose that $\mid 1 \ \rangle$ lies energetically higher than $\mid 2 \ \rangle$. The two transitions $\mid 3 \ \rangle$ - $\mid 1 \ \rangle$ and $\mid 3 \ \rangle$ - $\mid 2 \ \rangle$ with transition frequencies $\omega_1$ and $\omega_2$ are driven by two laser fields, $\omega_{l1}$ and $\omega_{l2}$, at Rabi frequencies $\epsilon_1$ and $\epsilon_2$, respectively. The atom can decay by spontaneous emission from $\mid 1 \ \rangle$ to $\mid 3 \ \rangle$ at a rate $2 \, \Gamma_1$ and from $\mid 2 \ \rangle$ to $\mid 3 \ \rangle$ at a rate $2 \, \nu$, where we neglect the decay channel $\mid 1 \ \rangle \rightarrow \mid 2 \ \rangle$ (although existing in In$^+$ at a rate $\Gamma_{^{3}P_{1} \rightarrow ^{3}P_{0}} \simeq 2 \pi \cdot$ 3.5 mHz). Because the lifetime  of the $\mid 2 \ \rangle$ state is so long, excitation of the $\mid 3 \ \rangle$ - $\mid 2 \ \rangle$ transition cannot be detected directly via scattered fluorescence photons. Instead, {\it excitation of the $\mid 2 \ \rangle$ state} is determined via {\it an interruption of fluorescence photons} on the fast $\mid 3 \ \rangle$ - $\mid 1 \ \rangle$ transition \cite{Dehmelt82}. The response of the system when tuning the laser $\omega_{l2}$ over the $\mid 3 \ \rangle$ - $\mid 2 \ \rangle$ resonance is thus probed by the amount of fluorescence photons scattered on the $\mid 3 \ \rangle$ - $\mid 1 \ \rangle$ - transition. In this aspect, the way of determining the {\it induced coherence significantly differs} from ususal detection techniques employed in $\Lambda$ systems where fluorescence is recorded on the same transition over which the probing laser is scanned.

We start the analysis from the master equation for the V-system:

\begin{eqnarray} \label{master1}
\dot \rho_{11} & = & -2 \Gamma_1 \rho_{11} + i \epsilon_1 \rho_{31} - i \epsilon_1^{*} \rho_{13}\\
\dot \rho_{12} & = & - ( \Gamma_1  + \nu + i ( \delta_1 - \delta_2 ) ) \rho_{12} + i \epsilon_1 \rho_{32} - i \epsilon^{*}_2 \rho_{13}\\
\dot \rho_{13} & = & - ( \Gamma_1 + \gamma_l + i \delta_1 ) \rho_{13} - i \epsilon_2 \rho_{12} - i \epsilon_1 ( \rho_{11} - \rho_{33})\\
\dot \rho_{22} & = & -2 \nu \rho_{22} + i \epsilon_2 \rho_{32} - i \epsilon^{*}_2 \rho_{23}\\
\dot \rho_{23} & = & - ( \nu + i \delta_2 + \gamma_l) \rho_{23} - i \epsilon_1 \rho_{21} - i \epsilon_2 ( \rho_{22} - \rho_{33} )\\
\rho_{33} & = & 1 - \rho_{11} - \rho_{22}
\end{eqnarray}

Here $\delta_i = \omega_{li} - \omega_i$ denotes the detuning between the laser and the atomic transition frequency for $i = 1,2$ and $\gamma_l$ is the linewidth of the clock laser. We include the linewidth of this laser as generally $\gamma_l$ can be much bigger than $\nu$. In steady state and for $\delta_1=0$ the variation of $\rho_{11}$ with $\delta_2$ is ploted in Fig. 1. For weak excitation of the $\mid 3 \ \rangle$ - $\mid 2 \ \rangle$ transition, the excitation spectrum reveals two separated, well resolved peaks (Fig. 1b). For these parameters, the weak beam acts as a probe which explores the distribution of population among the dressed states of the pumping transition from a third "spectator" level. The observed absorption spectrum obtained by tuning the probe beam over the atomic resonance leads thus to a double-peaked structure known as Autler-Townes splitting (ATS) \cite{Autler55}. 

If the power of laser $\omega_{l2}$ is increased the spectrum changes its form, to the one shown in Fig. 1c. In the limiting case where both lasers are strongly saturating the $\mid 3 \ \rangle$ - $\mid 1 \ \rangle$ - and $\mid 3 \ \rangle$ - $\mid 2 \ \rangle$ - transitions there appears within the broad resonance a sharp peak where $\rho_{11}$ increases almost to its initial value (obtained if laser $\omega_{l2}$ is absent or strongly detuned). When $\delta_2 = 0$ both lasers are in resonance, then a coherence between levels $\mid 1 \ \rangle$ and $\mid 2 \ \rangle$ is established by the presence of the clock laser which increases the emission of fluorescence photons on transition $\mid 3 \ \rangle$ - $\mid 1 \ \rangle$ and suppresses the absorption of laser $\omega_{l2}$ \cite{Fleischhauer92,Scully98}. It has been demonstrated previously that the strong pumping on the metastable transition could lead to stabilization of quantum fluctuations in the fluorescence spectrum \cite{Narducci90,Mossberg91}. Note that the increased fluorescence which appears in the center of an Autler-Townes splitting (case shown in Fig. 1b) can be clearly distinguished from the increase of fluorescence in case of the clock laser induced coherence (CIC) between levels $\mid 1 \ \rangle$ and $\mid 2 \ \rangle$ (Fig. 1c). In case of ATS with zero detuning, $\delta_1=0$, the excitation spectrum is given by two Lorentzians with equal width but blue- and red-shifted by an equal amount from the pump laser frequency $\omega_{l1}$. In case of CIC the excitation spectrum lineshape follows two Lorentzians of different widths and opposite amplitudes located at the same position $\delta_2 = 0$.

It is the purpose of this paper to investigate the behaviour of the system in these two limiting cases, and, in particular, to study the transition from one regime into the other. It will be shown that there exist well-defined conditions in which either one or the other behaviour is displayed. A sharp border line separates the two regimes. This border is determined by a single control parameter given by the ratio $\epsilon_1/\epsilon_2$ of the two Rabi-frequencies. This ratio determines via the common lower level the laser induced degree of coherence between the two upper states.

\section{Pump on Resonance} \label{zero}

In this section we assume that laser $\omega_{l1}$ is in resonance with the $\mid 3 \ \rangle$ - $\mid 1 \ \rangle$ transition, i.e. $\delta_1 = 0$. In what follows we employ the steady state solutions of the master equation and make use of the particular level scheme of In$^+$. Here, the two upper states of the three-level system have very different lifetimes, with a ratio of the order of $\Gamma_1/\nu \sim 10^5$. This means, we have:

\begin{equation}
\Gamma_1 >> \nu
\end{equation}

Assuming further that we use for the excitation of transition $\mid 3 \ \rangle$ - $\mid 2 \ \rangle$ a laser which is spectrally broader than $\nu$, say with a width $\gamma_l$, we still will have:

\begin{equation}
\Gamma_1 >> \gamma_l >> \nu
\end{equation}

Thus - even for strong saturation of the clock-transition $\mid 3 \ \rangle$ - $\mid 2 \ \rangle$ - we will work in the limit:

\begin{equation} \label{limit}
\Gamma_1, \epsilon_1 >> \gamma_l, \epsilon_2 >> \nu
\end{equation}

Under these conditions the steady state solutions of the master equation can be considerably simplified so that a clear understanding of the CIC is obtained. Using (\ref{limit}), we have proved that the expression for the population in the excited state $\mid 1 \ \rangle$ can be reduced to:

\begin{equation} \label{rho11}
\rho_{11} = \frac{\epsilon_{1}^{2}}{2 \epsilon_{1}^{2} + \Gamma_{1}^2} \cdot \left( 1 - \frac{\epsilon_{1}^{2} \epsilon_{2}^{2} \Gamma_1 (\delta_{2}^2 + \Gamma_{1}^2)}{(2 \epsilon_{1}^{2} + \Gamma_{1}^2) \nu} \cdot \frac{1}{(\delta_{2}^2 - \lambda_{1}^2)(\delta_{2}^2 - \lambda_{2}^2)} \right),
\end{equation}

where the $\lambda_i$ are given by

\begin{equation} \label{lambda}
\lambda_{1 , 2} = \sqrt{a \pm b},
\end{equation}

with

\begin{eqnarray}
a  &  =  &  \frac{(4 \epsilon_1^4 - \Gamma_1^4) \nu - \epsilon_1^2 \epsilon_2^2 \Gamma_1 }{2 ( 2 \epsilon_1^2 + \Gamma_1^2 ) \nu} \\
b  &  =  & \sqrt{ \frac{ ( \epsilon_{1}^2 \epsilon_{2}^2 \Gamma_1 - ( 2 \epsilon_1^2 + \Gamma_1^2 ) \nu \eta_ 1 ) ( \epsilon_{1}^2 \epsilon_{2}^2 \Gamma_1 - ( 2 \epsilon_1^2 + \Gamma_1^2 ) \nu \eta_ 2 )}{ ( 2 ( 2 \epsilon_1^2 + \Gamma_1^2 ) \nu )^2}}
\end{eqnarray}

and

\begin{equation} \label{eta}
\eta_{1 , 2} = 3 ( 2 \epsilon_1^2 + \Gamma_1^2 ) \pm 2 \sqrt{9 \epsilon_1^4 + 10 \epsilon_1^2 \Gamma_1^2 + 2 \Gamma_1^4}.
\end{equation}

Note that expression (\ref{rho11}) is independent of the laser linewidth $\gamma_l$. This is a consequence of the limit (\ref{limit}): terms in $\gamma_l$ appearing in sums with $\epsilon_1$, $\Gamma_1$ are neglected in the derivation of (\ref{rho11}). This means that a frequency-unstable clock laser does not deform the lineshapes predicted by Eq. (\ref{rho11}).

Since $a$ is real for arbitrary $\epsilon_1$, $\epsilon_2$, one can see from Eq. (\ref{rho11}) and (\ref{lambda}) that the bahaviour of the system depends critically on the value of $b$. If $b$ is purely imaginary, we have:

\begin{eqnarray}
\lambda_{1,2} & = & \sqrt{ a \pm i \vert b \vert }
\end{eqnarray}

that is:

\begin{equation} \label{lambda1}
\lambda_1 = \lambda_2^{*} = \lambda_0 + i \Gamma_0
\end{equation} 

so that:

\begin{equation} \label{rho11a}
\rho_{11} = \frac{\epsilon_{1}^{2}}{2 \epsilon_{1}^{2} + \Gamma_{1}^2} \cdot \left( 1 - \frac{\epsilon_{1}^{2} \epsilon_{2}^{2} \Gamma_1 ( \delta_{2}^2 + \Gamma_{1}^2 )}{( 2 \epsilon_{1}^{2} + \Gamma_{1}^2 ) \nu} \cdot \frac{1}{ [ ( \delta_{2} - \lambda_0 )^2 + \Gamma_0^2 ] \cdot [ ( \delta_{2} + \lambda_0 )^2 + \Gamma_0^2 ] } \right)
\end{equation}

Thus, for purely imaginary values of $b$, the excitation spectrum of the $\mid 3 \ \rangle$ - $\mid 1 \ \rangle$ transition as a function of $\delta_2$ is described by two Lorentzians with equal widths, positioned at $ \pm \lambda_0$. We here recover the well known Autler - Townes doublet. In the limit (\ref{limit}), the approximate position of the resonances is found to be:

\begin{equation} \label{posATS}
\lambda_0 = \epsilon_1 \sqrt{ \sqrt{1 + 2 \left( \frac{\Gamma_1}{\epsilon_1} \right)^2 + \left( \frac{\epsilon_2}{\epsilon_1} \right)^2 \frac{\Gamma_1}{\nu} } - \frac{\gamma_l}{\Gamma_1} - \left( \frac{\Gamma_1}{\epsilon_1} \right)^2 }
\end{equation}

Note the dependence of (\ref{posATS}) on the power of the clock laser. 

We next consider the possibility of $b$ being real. In this case calculations show that the following relation holds:

\begin{equation}
a \pm b < 0
\end{equation}

We therefore have:

\begin{equation}
\lambda_1 = i \sqrt{\vert a + b \vert} 
\end{equation}

\begin{equation}
\lambda_2 = i \sqrt{\vert a - b \vert} 
\end{equation}

In this case (\ref{rho11}) can be written as:

\begin{equation} \label{rho11b}
\rho_{11} = \frac{\epsilon_{1}^{2}}{2 \epsilon_{1}^{2} + \Gamma_{1}^2} \cdot \left( 1 - \frac{\epsilon_{1}^{2} \epsilon_{2}^{2} \Gamma_1 (\delta_{2}^2 + \Gamma_{1}^2)}{(2 \epsilon_{1}^{2} + \Gamma_{1}^2) \nu} \cdot \frac{1}{( \delta_{2}^2 + \vert \lambda_{1} \vert^2 ) \cdot ( \delta_{2}^2 +  \vert \lambda_{2} \vert^2 )} \right)
\end{equation}

We again find two Lorentzians in the spectrum, however located at line center $\delta_2 = 0$. The two peaks have now different widths, $\vert \lambda_1 \vert$ and $\vert \lambda_2 \vert$, respectively.

The real and imaginary parts of $\lambda_{1 , 2}$ are plotted as a function of $\epsilon_2$ in Fig. 2. As can be seen from the plots, there exists a critical value $\epsilon_{2}^{\, c}$ above which the real part of $\lambda_{1 , 2}$ vanishes and the imaginary parts start to differ. Below $\epsilon_{2}^{\, c}$ the reverse is true, i.e. the position of the resonances differ (equal in modulus but different in sign) and the linewidths become degenerate. A clear borderline separates the two regimes - determined by the condition $b = 0$ - which can be deduced from (13) to be:

\begin{equation} \label{border}
\epsilon_{2}^{\, c} = \epsilon_1 \cdot \sqrt{\left( 2 + \left( \frac{\Gamma_{1}}{\epsilon_{1}} \right)^2 \right) \left( 6 + 3 \left( \frac{\Gamma_1}{\epsilon_1} \right)^2 + 2 \sqrt{9 + 10 \left( \frac{\Gamma_1}{\epsilon_1} \right)^2 + 2 \left( \frac{\Gamma_1}{\epsilon_1} \right)^4} \right) \frac{\nu}{\Gamma_1}}
\end{equation}
 
The dependence of $\epsilon_{2}^{\, c}$ on $\epsilon_1$ is plotted in Fig. 3. As can be seen from (\ref{border}), the border critically depends on both, saturation $\frac{\epsilon_1}{\Gamma_1}$ and the ratio of the natural linewidths of the two upper states, $\frac{\Gamma_1}{\nu}$. For strong saturation $\epsilon_1 >> \Gamma_1$ the dependence (\ref{border}) simplifies to a linear one:

\begin{equation} \label{bordersimple}
\epsilon_{2}^{\, c} = \epsilon_1 \cdot \sqrt{24} \sqrt{\frac{\nu}{\Gamma_1}}
\end{equation}

Note that the position of the CIC resonance corresponds precisely to the line center of the broader lorentzian curve. The two lorentzians are thus superposed at the same detuning (at $\delta_2 = 0$) but have opposite signs. To see this more clearly another expression for $\rho_{11}$ can be derived which lends itself to describe in a more transparent way the behaviour of the system for $\epsilon_2 > \epsilon_{2}^{\, c}$. It is obtained by first taking the limit $\nu \rightarrow 0$ (well suited for the CIC regime where $\epsilon_2$ is supposed to be much larger than $\nu$) and then adding the remaining part of Eq. (\ref{rho11}) to the expression thus obtained. In this way one obtains:

\begin{equation} \label{rho11c}
\rho_{11} = \frac{\epsilon_{1}^2}{2 \epsilon_{1}^2 + \Gamma_{1}^2} + \frac{\epsilon_{1}^2}{\delta_{2}^2 + 2 ( \epsilon_{1}^2 + \Gamma_{1}^2)} - \frac{\epsilon_{1}^4 \epsilon_{2}^2 \Gamma_1}{(2 \epsilon_{1}^2 + \Gamma_{1}^2)^2 (\delta_{2}^2 + \frac{\epsilon_{1}^2 \epsilon_{2}^2 \Gamma_1}{(2 \epsilon_{1}^2 + \Gamma_{1}^2 ) \nu}) \nu}
\end{equation}

The first term in (\ref{rho11c}) is the fluorescence produced by the system if the clock laser were absent. It also gives the value of $\rho_{11}$ in the limit of very large detuning $\delta_2$. The next two terms in (\ref{rho11c}) are produced by a strong clock laser, although in the limit (\ref{limit}) the second term does not depend explicitly on the clock laser intensity. The width of this term is determined not only by the width of the transition $\mid 3 \ \rangle$ - $\mid 1 \ \rangle$ but also by the power of the laser on that transition. Looking at the last term in (\ref{rho11c}) we see that it becomes more and more broadened as we increase the intensity of the clock laser. Moreover, its width is determined by the natural linewidth of state $\mid 2 \ \rangle$. For large $\epsilon_1/\Gamma_1$, the ratio of the widths of the last two terms in (\ref{rho11c}) is approximately equal to:

\begin{equation}
\sqrt{6} \frac{\epsilon_2}{\epsilon_{2}^{\, c}}
\end{equation}

The reduction in fluorescence on the $\mid 3 \ \rangle$ - $\mid 1 \ \rangle$ transition described by the third term in (\ref{rho11c}) can be understood by the increase in population transfer to the $\mid 2 \ \rangle$ state as the intensity of the clock laser is pumped up. Therefore, clearly, the positive Lorentzian (second term in (\ref{rho11c})) has to be a coherence contribution \cite{Grynberg96}. To demonstrate this the behaviour of the coherence $\rho_{12}$ as a function of $\delta_2$ is plotted in this regime in Fig. 4.

\section{General Case - $\delta_1 \not= 0$} \label{general}

For the case of off-resonant pump, $\delta_1 \not= 0$, again analytical results can be obtained by solving the master equation. These, however, are cumbersome and hence only numerical results are presented. 

The overall behaviour of fluorescence on the fast transition in the two limiting cases $\epsilon_2 < \epsilon_{2}^{\, c}$ and $\epsilon_2 > \epsilon_{2}^{\, c}$ for $\delta_1 \not= 0$ is displayed in Fig. 5. For $\epsilon_2 < \epsilon_{2}^{\, c}$, the two peaks of the ATS regime can be identified, being asymmetric in this case (Fig. 5a). If $\epsilon_2 > \epsilon_{2}^{\, c}$ the CIC regime is entered. The CIC resonance is now shifted away from the atomic transition within the broad resonance of reduced fluorescence which stays approximately constant in frequency space (Fig. 5b). In order to understand this behaviour, we have to examine the poles of $\rho_{11}$ as a function of $\delta_2$. Let us denote the complex zeros of the denominator in $\rho_{11}$ by $\lambda_1$ and $\lambda_2$. We show in Fig. 6 the real and imaginary parts of $\lambda_i$, i = 1,2, as a function of $\delta_1$ in the two limiting cases of $\epsilon_2 < \epsilon_{2}^{\, c}$ and $\epsilon_2 > \epsilon_{2}^{\, c}$. As can be seen in Fig. 6a, asymmetries appear in the spectra in case of ATS. For $\delta_1 \not= 0$, the two resonances are shifted by an unequal amount towards (away) from the atomic resonance and the linewidths and amplitudes start to differ (see Fig. 5a): The peak moving towards the atomic resonance decreases in width and increases in amplitude where the opposite is true for the second peak.

In case of $\epsilon_2 > \epsilon_{2}^{\, c}$ the CIC resonance is blue-shifted for $\delta_1 > 0$ and red-shifted for $\delta_1 < 0$ (see Fig. 6b). In contrast, the broad lorentzian is not shifted much. It stays approximately constant in frequency space but rapidly reduces its width for increased detuning $\delta_1$.

\section{Conclusions} \label{concl}

In conclusion we investigated the response of a V-system with very different lifetimes of the two upper states when continously pumped on the fast transition and exposed to a laser of varying intensity on the slow transition. Two different regimes were discerned as a function of clock laser power in which the system behaves in fundamentally different ways. In the ATS regime the clock laser serves as a probe which senses the population of the light shifted dressed states of the fast transition. This leads to the well known double peak structure of the Autler-Townes splitting and corresponds to the pump-probe regime. In contrast, in the CIC regime, the clock laser, apart from transferring population to the $\mid 2 \ \rangle$ state, induces at the 2-photon resonance a coherence which suppresses the absorption of the clock laser and increases the emission of fluorescence photons on the fast transition. For $\delta_1 = 0$, transparent analytical formulas are obtained which describe in simple terms the lineshape, position and widths of the corresponding resonances. Moreover, a sharp borderline has been derived which accurately separates the two regimes. This borderline relates the two laser intensities and depends only on the saturation of the fast transition and the ratio of the two lifetimes involved. In case of strong saturation of the fast transition, a simple linear dependence is obtained (Eq. \ref{bordersimple}). Thus, for a given cooling laser intensity, the observed regime can be immediately derived from the clock laser intensity. Presently, in our lab, experiments are under way to investigate this behaviour in a single trapped In$^{+}$ ion \cite{Becker00a}. The lifetime ratio of the two upper states of $\sim 10^5$ in combination with the recently improved clock laser power \cite{Eichenseer00} makes In$^+$ an attractive candidate for such kinds of experiments.

\newpage

{\large \textbf{Figure Captions}}

\vspace{1cm}

{\bf Fig. 1}: (a) V-system in In$^+$: the $5s^{2 \enspace 1}S_{0}$ groundstate is coherently coupled to the two lowest lying excited states $5s5p \ ^{3}P_{1}$ and $5s5p \ ^{3}P_{0}$. Upper state decay rates are $2 \, \Gamma_1 = 2 \pi \cdot$ 360 kHz and $2 \nu = 2 \pi \cdot$ 0.82 Hz, respectively. (b) Autler-Townes-splitting regime (ATS): for weak clock laser power and $\delta_1=0$ the excitation spectrum of the fast transition reveals two separated, well resolved peaks of equal amplitude. (c) Regime of clock-laser induced coherence (CIC): for strong clock laser power and $\delta_1=0$ the emission of fluorescence photons on the fast transition is increased (absorption of the clock laser is suppressed) at $\delta_2=0$. (d) 3D-plot showing the population of level $\mid 1 \ \rangle$ for the change between the ATS regime to the CIC-regime as a function of cooling laser Rabi frequency $\epsilon_1$; parameters are $\delta_1=0$ and $\epsilon_2 = 0.1 \cdot \Gamma_1$.

\vspace{1cm}

{\bf Fig. 2}: Real (a) and imaginary (b) parts of $\lambda_{1,2}$ (in units of $\Gamma_1$) as a function of clock laser Rabi frequency $\epsilon_2$ for $\delta_1 = 0$ and $\epsilon_1 = 2 \cdot \Gamma_1$ (see Eq. (\ref{lambda})). At $\epsilon_2 = \epsilon_{2}^{\, c}$ the real part vanishes and the imaginary parts start to differ. 

\vspace{1cm}

{\bf Fig. 3}: The critical value of the clock laser Rabi frequency $\epsilon_{2}^{\, c}$ as a function of cooling laser Rabi frequency $\epsilon_1$. The border separating the two regimes of ATS and CIC is determined by $b = 0$ (see Eq. (10) - (13)).

\vspace{1cm}

{\bf Fig. 4}: Real part of $\rho_{12}$ as a function of $\delta_2$ in the CIC regime. The coherence $\rho_{12}$ is prominent only in this regime. Parameters are: $\epsilon_1 = 0.8 \cdot \Gamma_1$, $\epsilon_2 = 0.2 \cdot \Gamma_1$, $\delta_1 = 0$.

\vspace{1cm}

{\bf Fig. 5}: Fluorescence on the fast transition as a function of $\delta_2$ in the two limiting cases $\epsilon_2 < \epsilon_{2}^{\, c}$ (a) and $\epsilon_2 > \epsilon_{2}^{\, c}$ (b) for $\delta_1 = 7 \cdot \Gamma_1$. In the ATS regime two asymmetric peaks can be identified. In the CIC regime the 2-photon resonance is now shifted from the atomic resonance. Parameters are: in the ATS regime $\epsilon_1 = 10 \cdot \Gamma_1$, $\epsilon_2 = 10^{-3} \cdot \Gamma_1$, in the CIC regime $\epsilon_1 = 0.8 \cdot \Gamma_1$, $\epsilon_2 = 0.2 \cdot \Gamma_1$.

\vspace{1cm}

{\bf Fig. 6}: Real and imaginary part of $\lambda_{1,2}$ (in units of $\Gamma_1$) as a function of $\delta_1$ in (a) the ATS and (b) the CIC regime. Parameters for $\epsilon_1$ and $\epsilon_2$ are the same as in Fig. 5.


\newpage



\begin{thebibliography}{99}

\bibitem{Alzetta76}
G. Alzetta, A. Gozzini, L. Moi, G. Orriols, {\it Nuovo Cim.} {\bf B 36}, 5, (1976).

\bibitem{Gray78}
H. R. Gray, R. M. Whitley, C. R. Stroud Jr., {\it Opt. Lett.} {\bf 3}, 218, (1978).

\bibitem{Arimondo76}
E. Arimondo, G. Orriols, {\it Nuovo Cim.} {\bf 17}, 333, (1976).

\bibitem{Arimondo96}
E. Arimondo, in: {\em Progress in Optics XXXV}, Ed.: E. Wolf, {\it Elsevier Science, Amsterdam}, p. 257, (1996), and references therein.

\bibitem{Harris91}
K.-J. Boller, A. Imamoglu, S. E. Harris, {\it Phys. Rev. Lett.} {\bf 66}, 2593, (1991).

\bibitem{Scully91}
M. O. Scully, {\it Phys. Rev. Lett.} {\bf 67}, 1855, (1991).

\bibitem{Harris89}
A. Imamoglu, S. E. Harris, {\it Opt. Lett.} {\bf 14}, 1344, (1989).

\bibitem{Lange93}
A. Nottelmann, C. Peters, W. Lange, {\it Phys. Rev. Lett.} {\bf 70}, 1783, (1993).

\bibitem{Zibrov95}
A. S. Zibrov,M. D. Lukin, D. E. Nikonov, L. Hollberg, M. O. Scully, V. L. Velichansky, H. G. Robinson, {\it Phys. Rev. Lett.} {\bf 75}, 1499, (1995).

\bibitem{Padmabandu96}
G. G. Padmabandu, G. R. Welch, I. N. Shubin, E. S. Fry, D. E. Nikonov, M. D. Lukin, M. O. Scully,  {\it Phys. Rev. Lett.} {\bf 76}, 2053, (1996).

\bibitem{Hau99}
L. V. Hau, S. E. Harris, Z. Dutton, C. H. Behroozi, {\it Nature} {\bf 397}, 594, (1999).

\bibitem{Scully99}
M. M. Kash, V. A. Sautenkov, A. S. Zibrov, L. Hollberg, G. R. Welch, M. D. Lukin, Y. Rostovtsev, E. S. Fry, M. O. Scully, {\it Phys. Rev. Lett.} {\bf 82}, 5229, (1999).

\bibitem{Becker00}
Th. Becker, J. von Zanthier, A. Yu. Nevsky, Ch. Schwedes, M. N. Skvortsov, H. Walther, E. Peik, to be published;\\
J. von Zanthier, J. Abel, Th. Becker, M. Fries, E. Peik, H. Walther, R. Holzwarth, J. Reichert, Th. Udem, T. W. H\"ansch, A. Yu. Nevsky, M. N. Skvortsov, S. N. Bagayev,
{\it Opt. Comm.\/} {\bf 166}, 57 (1999);\\
E. Peik, J. Abel, Th. Becker, J. von Zanthier, H. Walther, {\it Phys. Rev.} {\bf A 60 }, 439, (1999).

\bibitem{Stenholm86}
S. Stenholm, {\it Rev. Mod. Phys.} {\bf 58 }, 699, (1986).

\bibitem{Toschek81}
P. E. Toschek, W. Neuhauser, in: {\em Atomic Physics 7\/}, Eds.: D. Kleppner, F. M. Pipkin, Plenum, New York, (1981).

\bibitem{Dehmelt85}
G. Janik, W. Nagourney, H. Dehmelt, {\it J. Opt. Soc. Am.} {\bf B 2}, 1251, (1985).

\bibitem{Blatt89}
M. Schubert, I. Siemers, R. Blatt, {\it Phys. Rev.} {\bf A 39}, 5098, (1989).

\bibitem{Toschek92}
I. Siemers, M. Schubert, R. Blatt, W. Neuhauser, P. E. Toschek, {\it Europhys. Lett.} {\bf 18}, 139, (1992).

\bibitem{Dehmelt82}
H. G. Dehmelt, IEEE Trans. Instr. Meas. {\bf 31}, 83, (1982).

\bibitem{Autler55}
S. H. Autler, C. H. Townes, {\it Phys. Rev.} {\bf 100}, 703, (1955);\\
R. M. Whitley, C. R. Stroud, Jr., {\it Phys. Rev. A} {\bf 14}, 1498, (1976);\\
S. Stenholm, {\em Foundations of Laser Spectroscopy\/}, Wiley, New York (1984).

\bibitem{Fleischhauer92}
M. Fleischhauer, C. H. Keitel, M. O. Scully, Chang Su, B. T. Ulrich, and Shi-Yao Zhu {\it Phys. Rev. A} {\bf 46}, 1468, (1992).

\bibitem{Scully98}
G. R. Welch, G. G. Padmabandu, E. S. Fry, M. D. Lukin, D. E. Nikonov, F. Sander, M. O. Scully, A. Weis, F. K. Tittel, {\em Foundations of Physics\/}, {\bf 28}, 621, (1998).

\bibitem{Narducci90}
L. M. Narducci, G.-L. Oppo, M. O. Scully, Opt. Comm. {\bf 75}, 111, (1990).

\bibitem{Mossberg91}
D. J. Gauthier, Yifu Zhu, T. W. Mossberg, {\it Phys. Rev. Lett.} {\bf 66}, 2460, (1991).

\bibitem{Grynberg96}
G. Grynberg, M. Pinard, P. Mandel, {\it Phys. Rev. A.} {\bf 54}, 776, (1996).

\bibitem{Becker00a}
Th. Becker, M. Eichenseer, A. Yu. Nevsky, E. Peik, T. Schneider, Ch. Schwedes, C. Skornia, J. von Zanthier, H. Walther, work in progress.

\bibitem{Eichenseer00}
M. Eichenseer, A. Yu. Nevsky, E. Peik, J. von Zanthier, H. Walther, to be published.


\end{thebibliography}
\end{document}